\documentclass[ aps, prl, twocolumn, 10pt,showpacs,nofootinbib,superscriptaddress,longbibliography]{revtex4-2}
\usepackage{graphicx}
\usepackage{amsfonts}
\usepackage{amsmath}
\usepackage{bm}
\usepackage{hyperref}

\begin{document}
\title{Orbital angular momentum can take non-integer values in a closed universe}
\author{Daniel Burgarth}
\affiliation{Department Physik, Friedrich-Alexander-Universit\"at Erlangen-N\"urnberg, Staudtstraße 7, 91058 Erlangen, Germany} 
\author{Paolo Facchi}
\affiliation{Dipartimento di Fisica, Universit\`a di Bari, I-70126 Bari, Italy}
\affiliation{INFN, Sezione di Bari, I-70126 Bari, Italy}
\date{\today}
\begin{abstract}
We show that the spectrum of orbital angular momentum in quantum mechanics consists of two parts
when the underlying space has periodic boundaries. While the first part
consists of the usual textbook integer quantized values, the second is a continuous band arising from regions at the `edge' of space with respect to the center of rotation.  The spectrum thus contains not only half-integer values, previously thought impossible for orbital angular momentum, but even irrational ones. Remarkably, this effect is independent of the size of space. While these spectral components remain undetectable in laboratory experiments, they could still produce observable effects on cosmological scales, for instance in the Cosmic Microwave Background Radiation.
\end{abstract}

\maketitle
\emph{Introduction:--} Orbital angular momentum $\vec{L} =\vec{r}\times \vec{p}$ has played a central role in the development of quantum mechanics in the early twentieth century~\cite{heisenberg_uber_1925}. Its integer quantization in Heisenberg's new quantum mechanics was already proven in 1926 by Lucy Mensing~\cite{mensing_rotations-schwingungsbanden_1926}. Nevertheless, the question of quantization has been revisited many times since, see~\cite{biedenharn_angular_1981} for a stimulating review. The usual textbook reasoning through ladder operators and their commutator relationships~\cite{sakurai_modern_2021} only constrains the potential eigenvalues of the component $L_z$ of orbital angular momentum to integer \emph{and} half-integer values. Because half-integer values are also realized in nature by spin degrees of freedom, the key question was why half-integer values are not found in the spectrum of orbital angular momentum. Of course, once orbital angular momentum is defined on the usual Hilbert space of square integrable wavefunctions, $L^2(\mathbb{R}^3)$, it is an easy exercise to show that the possible values of $L_z$ are integer multiples of the reduced Planck constant. However, the quest for good fundamental reasons of why the Hilbert space might not be something more exotic, like a Riemann sheet (often referred to as `multi-valued' wavefunctions), plagued some, including Wolfgang Pauli, who visited the subject three times~\cite{bethe_quantentheorie_1933, pauli,flugge_encyclopediaof_1958} (see also~\cite{merzbacher_single_1962}). Therefore, it was suggested that purely algebraic proofs, based soley on canonical commutation relationships, are superior, since they would not depend on the underlying Hilbert space~\cite{Buchdahl2005,ballentine_quantum_2015}.

Here we show that, surprisingly, there \emph{are} Hilbert spaces in which an orbital angular momentum $L_z=x p_y - y p_x$ made of \emph{canonical} position and momentum operators can have not only half-integer elements in the spectrum, but even irrational ones. Despite this, the wavefunction is `single valued'. This was overlooked in the algebraic arguments because these values are in the continuous spectrum and do not have normalizable eigenstates. And, even more surprisingly, such a Hilbert space might well be ours, namely a closed universe.

\begin{figure}[ht]
     \centering
     \includegraphics[width=0.9\columnwidth,
      trim=1.5cm 2.2cm 1.5cm 3.5cm,
     clip
     ]{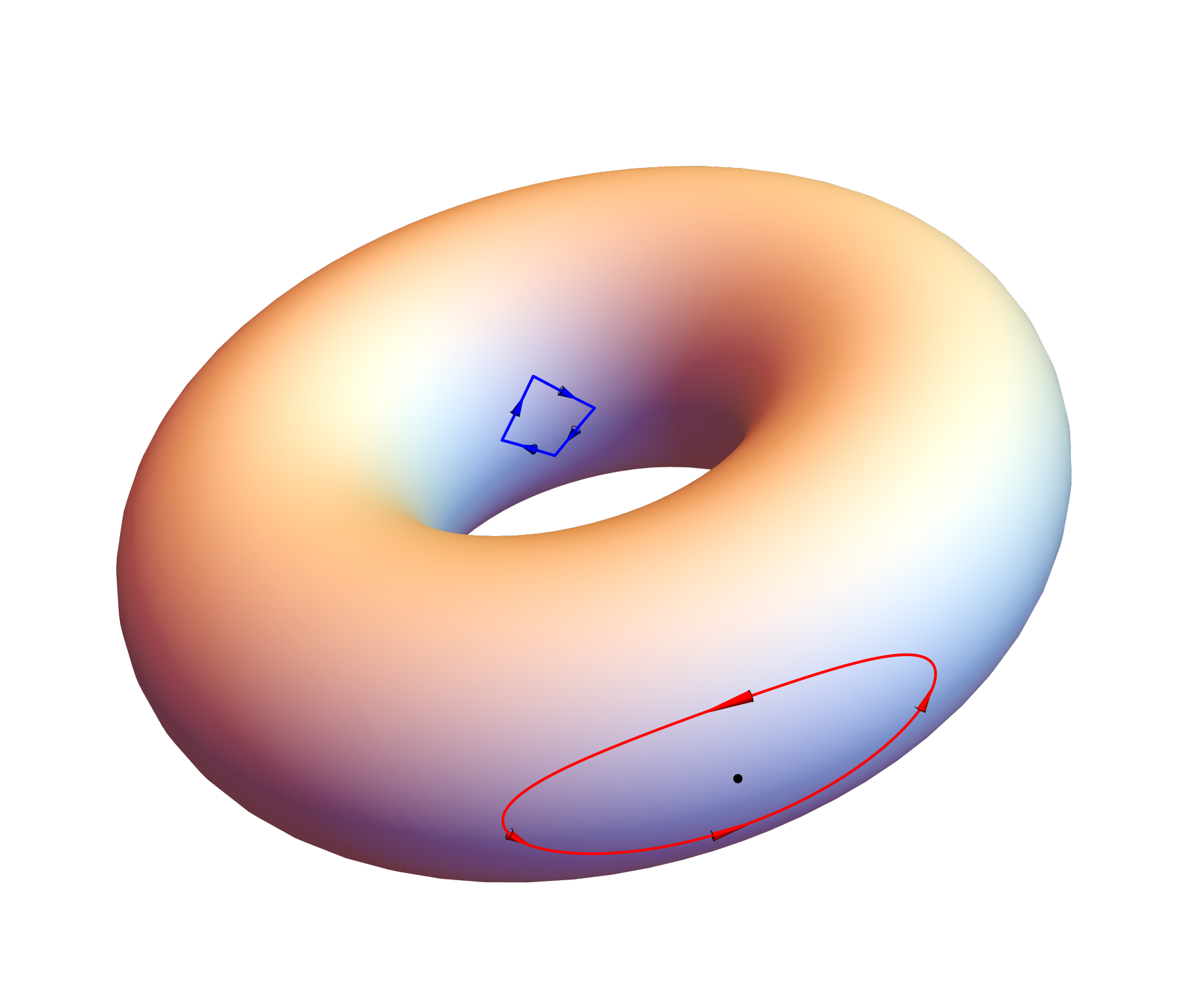}
     \caption{Although we are considering the \emph{flat} torus, it is interesting to visualize the trajectories on a torus embedded non-isometrically in 3D. Here, we can see the usual rotation (red) nearby the center of rotation (black dot) as well as the edge state (blue) far from the rotation center which leads to the continuous spectrum of $L_z$. Observe the inverted chirality of the blue orbit which appears to rotate around a different center.}
     \label{fig:torus3d}
\end{figure} 

In fact, we will show that in a closed universe there are states which have \emph{arbitrary} spectral values of $L_z$ with modulus larger than $\hbar$. These states need to be rotated by generally irrational angles to return to themselves, and thereby describe anyons~\cite{anyons}. 

Remarkably, neither the spectrum of angular momentum nor its square, as a measure of energy, depends on the size of the closed universe: this is not a local effect, but is caused by the global topology given by periodic boundary conditions. The simple reason is that orbital angular momentum $L_z$, which is made of products of multiplication and derivative operators, is invariant under scale transformations. This is a crucial difference to other potential topological fingerprints considered in the literature, which disappear in a large universe~\cite{paraskevas_probing_2025}. 

\begin{figure*}[ht]
     \centering
     \includegraphics[width=0.9\textwidth]{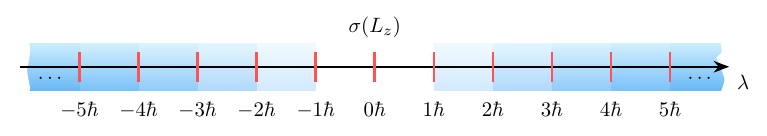}
     \caption{Schematics of the spectrum of orbital angular momentum $L_z$ in a closed universe. Red lines indicate the usual integer eigenvalues. The blue continuum arises from the periodic boundary conditions and is independent of the diameter $\sqrt2\ell$ of space. The red lines are infinitely degenerate, whereas the degeneracy of the $d$-th band counting from the origin is $d$ and its increase is indicated by the color saturation of the bands. There is a band-gap in the interval $(-
     \hbar, \hbar)$. The height of the lines and bands is for readability and has no physical meaning.}
     \label{fig:spectrum}
\end{figure*} 

The wavefunctions of the exotic spectrum are placed far away from the respective center of rotation of the angular momentum (see Fig.\ref{fig:torus3d}). That means that they cannot be prepared or measured in the lab, where we observe only the standard values of the angular momentum. In this sense, the result is fully consistent with our quantum experiments.

However, they could have been populated in the early universe and give a contribution to the energy of the cosmos. 
Further, the new states exhibit a chirality opposite to the usual one and, surprisingly, rotate in the opposite direction than expected around a different apparent center of rotation (see Fig.~\ref{fig:torus3d}). It is therefore possible that such states have an impact on observable quantities, such as the cosmic microwave background radiation (CMBR). Recent data \cite{cmbr} examines the possibility of a torodial flat universe and it would be very interesting to re-examine this in the context of the expected spatial profile of orbital edge states. Another option would be to consider the orbital angular momentum of the photon which might lead to observable consequences in the CMBR \cite{cmbr1}. 

Interestingly, the same effect occurs on the small scale in compactified dimensions such as those considered in string theory. We remark that here, when compactifying free space, the purely integer spectrum of orbital angular momentum changes to a continuous one. This is a surprising topological effect and the opposite of what one would expect from confinement of a quantum system.

On a mathematical level, the representation of angular momentum on the closed universe is no longer integrable in the precise sense of~\cite{schmudgen_invitation_2020}. This means that commutation relations still hold on the level of generators, but on the physical level of evolutions, new opportunities arise, namely the anyons. We were able to find the full analytical spectral representation of orbital angular momentum for the closed universe and for compactified dimensions.

\emph{Intuition:--}
Our key result is that orbital angular momentum in a closed universe, in addition to the ordinary integer eigenvalues $\hbar \mathbb{Z}$, also has an exotic continuous spectrum on $\mathbb{R}\setminus (-\hbar,\hbar)$ (see  Fig.~\ref{fig:spectrum}). While we sketch a complete technical proof below, it is instructive to give a semiclassical argument first, which gives the correct result. Here, we will consider the component $L_z$ of angular momentum and not $|\vec{L}|^2$ which is considerably harder and where we only have partial results till date. 

Since $L_z$ is acting on the $xy$-plane only, it suffices to restrict oneself to two spatial dimensions. We will consider a flat finite universe with periodic boundary conditions and dimension $2\ell \times 2\ell$. We found that other aspect ratios lead to chaotic trajectories which will be discussed elsewhere, since they are not directly related to our key result. 

\begin{figure}[ht]
     \centering
     \includegraphics[width=0.9\columnwidth]{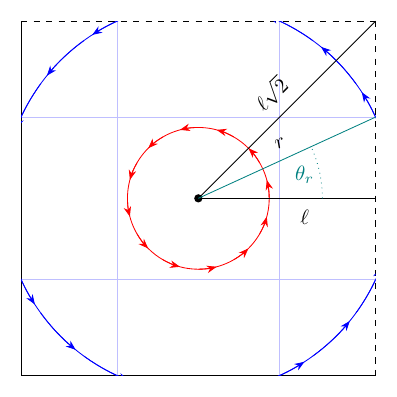}
     \caption{Classical rotations around the center of a flat torus. Dashed lines correspond to edges which are identified with their opposing ones through the periodic boundary condition. For a small radius $r<\ell$ we have a usual circle (red curve). For $\ell < r \le \ell \sqrt{2}$ the particle trajectory (blue curve) appears to jump (thin blue lines) although in reality, the trajectory is continuous due to the periodic boundary conditions. The allowed angles for the blue curve starts with $\theta_r=\arccos{\left(\frac{\ell}{r}\right)}$. The effective circumference $2\pi\mu(r)=2\pi \left(1-\frac{4}{\pi}\theta_r\right)$ of the blue trajectory is shorter than it would be due in open space due to such `shortcuts'.}
     \label{fig:torus}
\end{figure}

The main intuitive idea here is to consider the classical orbits and impose a Bohr-Sommerfeld quantization condition. This strategy, referred to as `sewing flesh on classical bones' in a seminal paper by Berry and Mount~\cite{berry_semiclassical_1972} is particularly effective when the original Hamiltonian is quadratic in position and momentum, as is the case here. Consider Fig.~\ref{fig:torus} for the two types of classical orbits one has on a flat torus: one is the usual circular one, whereas the other, close to the `edge' of the universe as seen from the center of rotation, is confined to a narrower range of angles, while still fulfilling the quantization condition. Such confinement leads to higher spectral values of $L_z$.  
More specifically, an integer number $m$ of wavelengths must fit along the whole orbit: $m\lambda=2\pi r \mu(r)$, where $\mu(r)$ is the ratio of the circumferences of the orbit to the full circle. 
 The momentum is related to the wavelength by the de Broglie relation $p=h/\lambda$, where $h$ is Planck's constant. Therefore, for an orbit of radius $r$ one has that $L_z = r p = \hbar m /\mu(r)$, with $\hbar=h/2\pi$.
 For any radius $r<\ell$ we obtain  ordinary circular orbits with $\mu(r)=1$, and the momentum $L_z=\hbar m$ is quantized. 
As we move further out towards the edge ($r>\ell$), the ratio $\mu (r)$ shrinks continuously to zero, implying a continuous spectral band for all integers $m\neq 0$. Traversing the orbits multiple times leads to increased degeneracy along the continuous spectrum. This argument is made rigorous by the fibered Hilbert space approach used in the next paragraph.

\emph{Full quantum proof:--} Here we provide a rigorous proof of the intuition described above. This is necessarily a bit technical, but the main idea is to solve the Schr\"odinger equation along the classical orbits as a function of their distance from the rotation center. The key references for the notation and mathematical physics background used are~\cite{hall_quantum_2013, reed_methods_2005}. To set things up, consider wavefunctions on a torus. The Hilbert space is given by
\begin{equation}
    \mathcal{H}= L^2(\mathbb{T}^2_{\ell})\simeq L^2(\mathbb{S}_{\ell/\pi})\otimes L^2(\mathbb{S}_{\ell/\pi}),
\end{equation}
where
$\mathbb{T}_{\ell}^2 = \mathbb{S}_{\ell/\pi}\times  \mathbb{S}_{\ell/\pi}$
is the flat 2-torus and  $\mathbb{S}_{r}$ is the circle of radius $r$, i.e., the boundary of the open ball $\mathbb{B}_r$.

The orbital angular momentum operator is then defined by
\begin{equation}
    L_z = x p_y - y p_x \simeq q\otimes p - p\otimes q,
    \label{eq:Lz}
\end{equation}
where the position operator $q=q^\dagger$ is the usual multiplication operator 
\begin{equation}
    (q\psi)(x) = x \psi (x), \quad \psi\in L^2(\mathbb{S}_{\ell/\pi})\simeq L^2([-\ell,\ell]).
\end{equation}
Due to the finite size, $\|q\|=\ell$ and $q$ is a bounded operator defined on all of $ L^2([-\ell,\ell])$. Notice that the maximal length in such universe is $\max d(\bm{x},\bm{y})=\sqrt{2}\ell$.

The momentum operator $p=p^\dagger$ is given by
\begin{equation}
    (p\psi)(x)= -i \hbar \frac{d}{dx} \psi(x),  \quad \psi\in H^1(\mathbb{S}_{\ell/\pi})
\end{equation}
where the domain is the first Sobolev space with periodic boundary conditions,
\begin{equation}
    H^1(\mathbb{S}_{\ell/\pi})\simeq \bigl\{\psi\in H^1([-\ell,\ell])\, : \, \psi(-\ell)=\psi(\ell)\bigr\}.
\end{equation}
 That is, the momentum is defined on periodic wavefunctions such that their (weak) derivative exists and is square integrable. 

Whenever defined, the canonical commutation relation $[q,p]=i \hbar$ holds: in this sense, $q$ and $p$ are canonical conjugate observables. They do not, however, exponentiate to the Weyl relationships~\cite{hall_quantum_2013}.
The spectrum of $q$ is given by
\begin{align}
    \sigma(q) &=\sigma_{\mathrm{ac}}(q)=[-\ell,\ell]\,,
    \label{eq:specp1}
\end{align}
where `$\mathrm{ac}$' stands for an absolutely continuous spectrum~\cite{hall_quantum_2013}. In particular, no normalizable (proper) eigenstates exist. The spectrum of $p$ is given by
\begin{align}
    \sigma(p) &=\sigma_{\mathrm{pp}}(p)=\frac{\pi\hbar}{\ell} \mathbb{Z}\, .
    \label{eq:specp2}
\end{align}
Here, `$\textrm{pp}$' stands for pure point spectrum~\cite{hall_quantum_2013}, and in particular, normalizable eigenfunctions exist. In fact the eigenfunctions of $p$, 
\begin{equation}
    u_n(x) = \frac{1}{\sqrt{2\ell}}\, e^{i \frac{n\pi}{\ell}x}, \quad n\in \mathbb{Z} ,
\end{equation}
form an orthonormal basis (the well-known Fourier basis on $[-\ell,\ell]$).
By the analytic vector theorem, it is easy to see that the angular momentum operator $L_z$ on $D(L_z)= C^{\infty}(\mathbb{T}^2_{\ell})$ is essentially self-adjoint.

The unusual spectral properties arise from the tension between rotations and the square determining the boundary conditions. In particular, by going to polar coordinates $x=r\cos(\phi)$, $y=r\sin(\phi)$ one gets that the range $\Gamma_r$ of the angle $\phi$ depends on the radius $r$:  for $r<\ell$, $\Gamma_r=\mathbb{S}_1$, while for $r\in[\ell, \sqrt{2}\ell]$, 
\begin{equation}
    \Gamma_{r}  
    =\bigcup_{j=0}^3 \left[j\frac{\pi}{2}+\theta_r,\; (j+1)\frac{\pi}{2}-\theta_r\right),
\end{equation}
with $\theta_r = \arccos(\ell/r)$. See Fig.~\ref{fig:torus}.

\begin{figure}[ht] 
     \centering
     \includegraphics[width=1.0\columnwidth,
     trim=2cm 0cm 2cm 0cm,
  clip
     ]{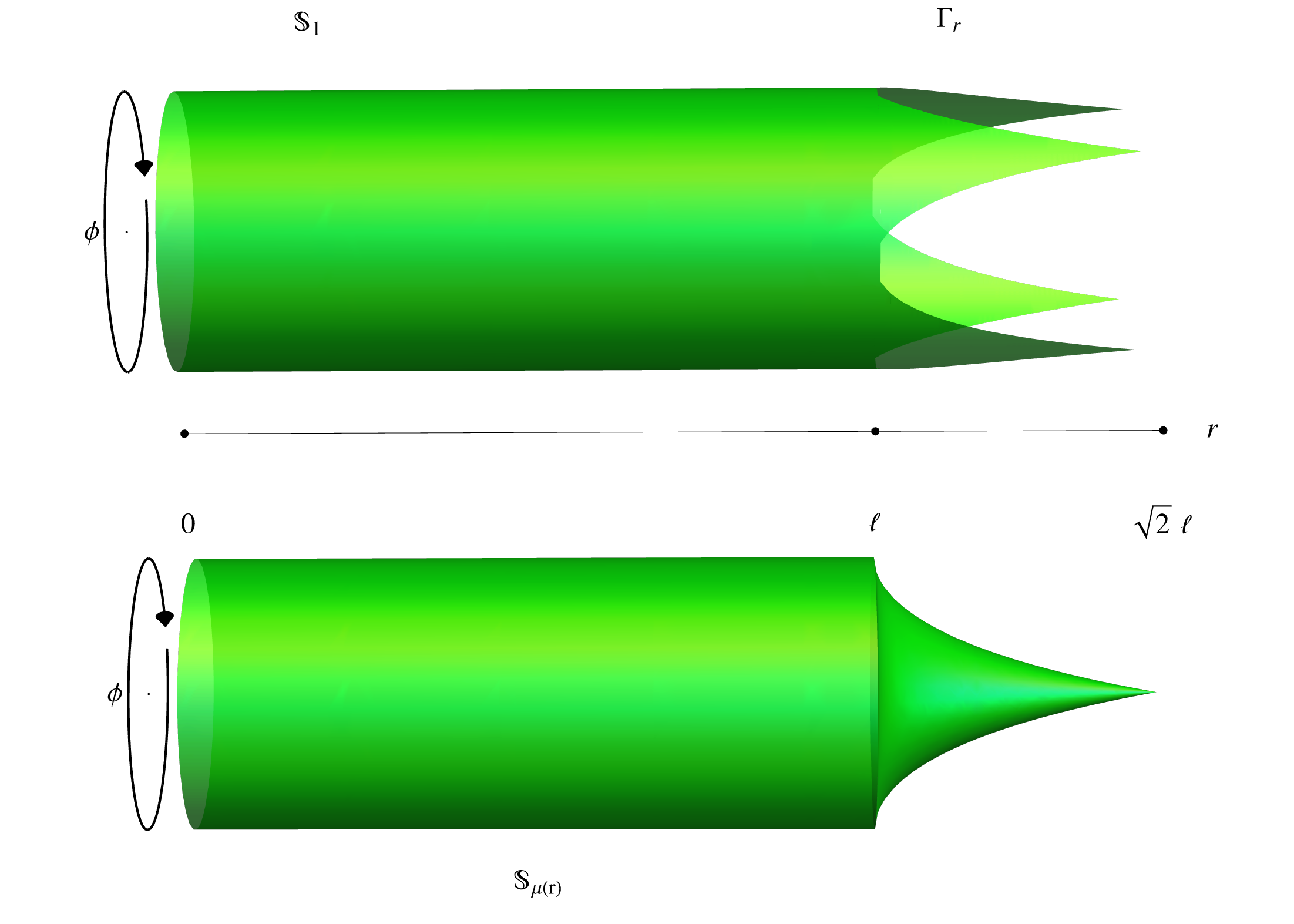}
     \caption{Fibration: the two-dimensional manifold $\mathbb{T}_\ell^2$ can be fibered by the classical orbits as a fiber bundle over the one-dimensional manifold $(0,\sqrt{2}\ell)$. In the upper picture, the fibers are isomorphic to circles of radius one for $r\le \ell $ and to the four disconnected line segments $\Gamma_r$, with consecutive edges identified, for $r>\ell$. In the lower picture, these four segments are identified with a circle of effective radius $\mu(r)$. The fiber bundle induces a fibration of the corresponding Hilbert spaces.}
     \label{fig:fibrebundles}
\end{figure} 

Motivated by Fig.~\ref{fig:fibrebundles} it is useful to split (fiber) the Hilbert space as a direct integral  (a continuous generalization of the direct sum). The advantage is that \emph{within} the fibers, we will find normalizable eigenstates even for the bands (the absolutely continuous spectrum) of $L_z$ and we will be able to fully characterize the spectrum without requiring many technicalities~\cite{reed_methods_2005}. We get the fibration over $r$
\begin{equation}
    \mathcal{H} \simeq \int_{[0,\sqrt{2}\ell]}^{\oplus} \mathcal{H}_r \,r dr, \qquad \mathcal{H}_r \simeq L^2(\Gamma_r),
\end{equation}
with the scalar product
\begin{align}
\langle \psi_1|\psi_2\rangle &= \int_{[0,\sqrt{2}\ell]} \langle \psi_1(r,\cdot)|\psi_2(r,\cdot)\rangle_{\mathcal{H}_r} r dr
\nonumber\\
&= \int_{[0,\sqrt{2}\ell]}\int_{\Gamma_r}\psi_1(r,\phi)^*\psi_2(r,\phi)\, r d\phi\, dr .
\end{align}
By using the fact that (see Fig.~\ref{fig:fibrebundles})
\begin{equation}
L^2(\Gamma_r)\simeq L^2(\mathbb{S}_{\mu(r)}) ,
\end{equation}
where
\begin{equation}
  \mu(r)=  \begin{cases}
			1, & \text{if $r\in[0,\ell)$}\\
            1-\frac{4}{\pi}\arccos{\left(\frac{\ell}{r}\right)}, & \text{if $r\in[\ell,\sqrt{2}\ell]$}
		 \end{cases}  ,
\end{equation}
 one has
\begin{equation}
    \mathcal{H} \simeq \int_{[0,\sqrt{2}\ell]}^{\oplus} \mathcal{H}'_r \,r dr, \qquad \mathcal{H}'_r \simeq L^2(\mathbb{S}_{\mu(r)}),
\end{equation}
and the angular momentum is represented as
\begin{align}
    {L_z} &\simeq \int_{[0,\sqrt{2}\ell]}^{\oplus} p_{\phi}(r) \, r dr, \\
    p_{\phi}(r) &= -i \hbar \frac{d}{d\phi}, \quad   D\bigl(p_{\phi}(r)\bigr) = H^1(\mathbb{S}_{\mu(r)}).
    \label{eq:pphi}
\end{align}

\begin{figure}[ht]
     \centering
     \includegraphics[width=0.9\columnwidth]{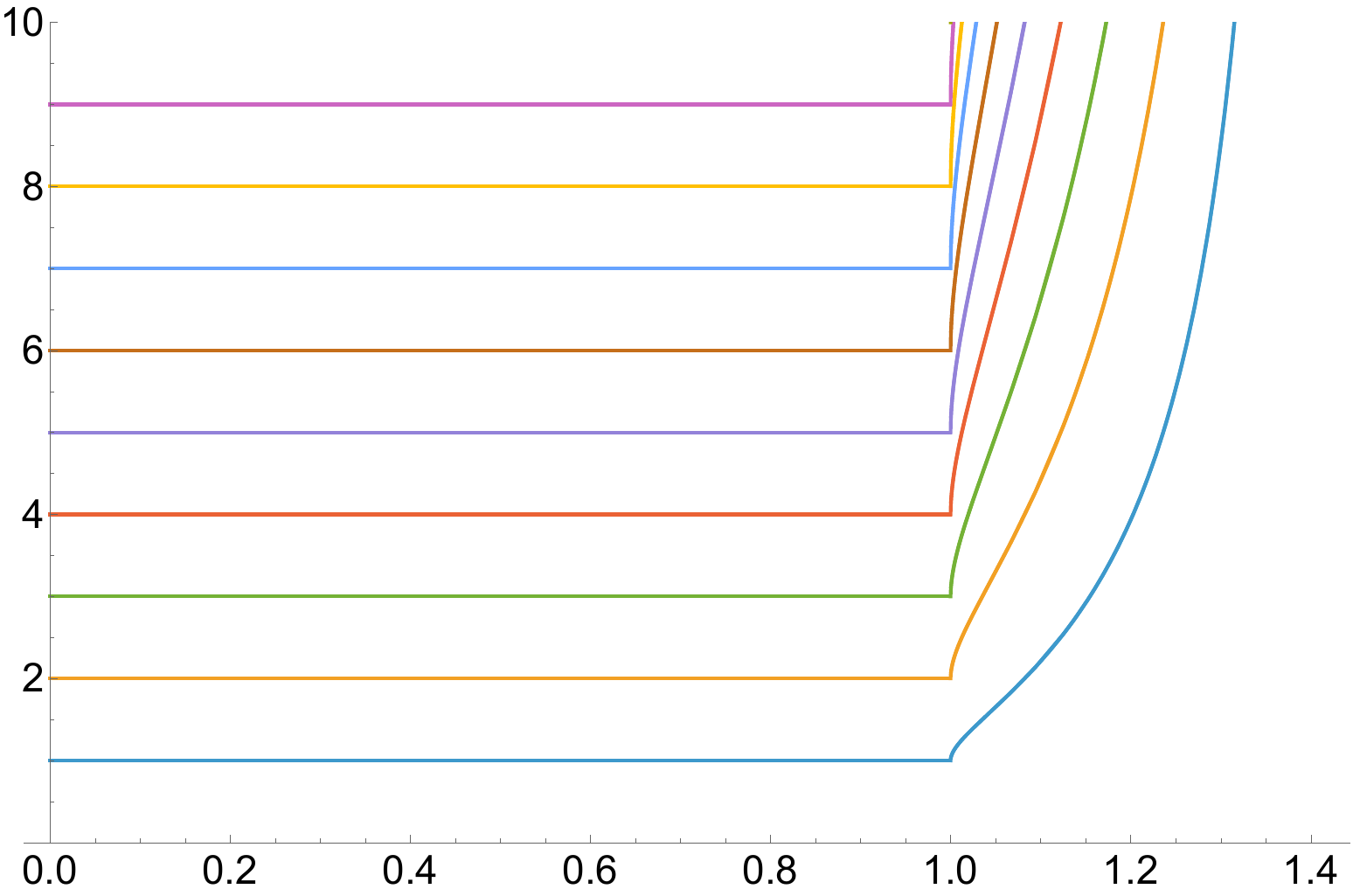}
     \caption{Eigenvalues $\lambda_m$ of the fiber operator $p_\phi(r)$ as a function of $r/\ell$}
     \label{fig:eigenvalues}
\end{figure}

For the spectrum of the fiber operator~\eqref{eq:pphi}, we have
from~\eqref{eq:specp2} with $\ell=\pi\mu(r)$ that
\begin{align}
    \sigma(p_{\phi}(r)) &=\sigma_{\mathrm{pp}}(p_{\phi}(r))=\frac{\hbar}{\mu(r)} \mathbb{Z},
\end{align}
and a corresponding orthonormal basis of eigenfunctions of $p_{\phi}(r)$ (Fourier basis on $[-\mu(r)\pi,\mu(r)\pi]$) is given by
\begin{equation}
    u_m(\phi) = \frac{1}{\sqrt{2 \pi\mu(r)}}\, e^{i \frac{m}{\mu(r)}\phi}, \quad m\in \mathbb{Z}.
\end{equation}
More explicitly the eigenvalues are given by
\begin{equation}
    \lambda_m(r) = \frac{\hbar m}{\mu(r)} = \begin{cases}
			\hbar m, & \text{if $r\in[0,\ell)$}\\
            \frac{\hbar m}{1-\frac{4}{\pi}\arccos{\left(\frac{\ell}{r}\right)}}, & \text{if $r\in[\ell,\sqrt{2}\ell)$}
		 \end{cases}, 
\end{equation}
for all $m\in\mathbb{Z}$. See Fig.~\ref{fig:eigenvalues}.

Finally, with the spectrum of the fiber operator, we can determine the spectrum of $L_z$ simply applying Theorem XIII.85 of~\cite{reed_methods_2005}. This theorem allows us to identify the spectrum simply by looking at Fig.~\ref{fig:eigenvalues}. Roughly speaking, the values with non-zero measure preimage on the $r$-axis are pure point eigenvalues, while the ones with zero measure preimage are part of the continuous spectrum. We observe that the spectrum splits into a pure point and an absolutely continuous spectrum as
\begin{equation}
    \sigma(L_z)=\sigma_{\mathrm{pp}}(L_z) \cup \sigma_{\mathrm{ac}}(L_z)\, . 
\end{equation}

The pure point spectrum is given by the usual textbook values
 \begin{equation}
     \sigma_{\mathrm{pp}}(L_z)= \hbar \mathbb{Z}
 \end{equation}
 with infinite degeneracy. The corresponding normalizable eigenfunctions $\in \mathcal{H}_{\mathrm{pp}}(L_z)$ are
\begin{equation}
    v_m(r,\phi) = w(r) u_m(\phi), \quad w\in L^2, \quad \operatorname{supp} (w)\subseteq [0,\ell],
\end{equation}
for $m\neq 0$, and $v_0(r,\phi)=w(r)$, with $w\in L^2$.
Notice that the eigenfunctions corresponding to the ordinary spectrum for all nontrivial rotation are supported in the closure of the disk~$\mathbb{B}_\ell$.

The absolutely continuous spectrum is given by
\begin{equation}
     \sigma_{\mathrm{ac}}(L_z)= (-\infty, -\hbar] \cup [+\hbar,+\infty)
 \end{equation}
with  finite degeneracy $d\in \mathbb{N}$, $d\geq1$  in the bands
 \begin{equation}
\bigl(-\hbar (d+1), -\hbar d\bigr] \quad \text{and}\quad \bigl[\hbar d, \hbar (d+1)\bigr).     
 \end{equation}
The corresponding absolutely continuous wavefunctions  
\begin{equation}
    \psi\in\mathcal{H}_{\mathrm{ac}}(L_z), \quad
    \quad \psi\in L^2, \quad \operatorname{supp} (\psi)\subseteq \mathbb{T}^2_{\ell}\setminus \mathbb{B}_{\ell}
\end{equation}
are supported near the corners $\ell < r <  \sqrt{2}\ell$, at the farthest edge of the universe with respect to the center of rotation.

\emph{Conclusion and outlook:--} Our results show that algebraic proofs are unable to exclude half-integer valued orbital angular momentum: in a closed flat universe, angular momentum has a strange spectrum consisting of the usual integer multiples of Planck's constant \emph{and} a surprising continuous component. Since the continuous part cannot be observed in the lab, corresponding to wavefunctions which are located far away, it is fully consistent with quantum experiments, but might have implications on a cosmic scale. For instance, the photon orbital angular momentum might be lead to predictions for the CMBR.  Our results equally apply to very small spaces with periodic boundary conditions and may therefore also be relevant for theories with compactified dimensions. It would furthermore be interesting to generalize our work to interacting particles, to curved spacetime and to theories of quantum gravity. Clearly, our complete spectral analysis of $L_z$ is only a small step toward implications in other fields, and we hope this work will stimulate a rich interdisciplinary discussion.

\emph{Acknowledgments:--}
PF acknowledges support from INFN through the project ``QUANTUM'', from the Italian National Group of Mathematical Physics (GNFM-INdAM), from PNRR MUR projects CN00000013-``Italian National Centre on HPC, Big Data and Quantum Computing'', and from the Italian funding within the ``Budget MUR - Dipartimenti di Eccellenza 2023--2027''  - Quantum Sensing
and Modelling for One-Health (QuaSiModO).

\bibliographystyle{prsty-title-hyperref}

\end{document}